\documentclass[]{aipproc}

\layoutstyle{6x9}

\usepackage{graphicx}
\usepackage{amssymb,bm,mathrsfs,bbm,amscd}
\usepackage[tbtags]{amsmath}

\begin{document}

\title{The role of hyperfine mixing in $b\to c$ semileptonic and
  electromagnetic decays of doubly-heavy baryons\thanks{Supported by
    DGI and FEDER funds, under contracts FIS2008-01143/FIS,
    FIS2006-03438, FPA2007-65748, CSD2007-00042, by Junta de Castilla
    y Le\'on under contracts SA016A07 and GR12, and by the EU
    HadronPhysics2 project}}


\author{C. Albertus}{address={Departamento de F\'{\i}sica Fundamental, Universidad de Salamanca, E-37008 Salamanca, Spain}
}

 \author{E. Hern\'andez}{address={Departamento de F\'{\i}sica
      Fundamental e IUFFyM, Universidad de Salamanca, E-37008
      Salamanca, Spain}
} 

\author{J. Nieves}{address={Instituto de
    F\'{\i}sica Corpuscular (IFIC), Centro Mixto CSIC-Universidad de
    Valencia, Institutos de Investigaci\'on de Paterna, Aptd. 22085,
    E-46071 Valencia, Spain}
}

\begin{abstract}
We analyze the effects of hyperfine mixing in $b\to c\, $ semileptonic
and electromagnetic decays of doubly heavy baryons.
\end{abstract}

\maketitle

\section{Introduction}
In the infinite heavy quark mass
limit, and according to heavy quark spin symmetry, the total spin of the 
heavy quark subsystem in a doubly heavy baryon
can be selected to be $S_h=0,1$.  Indeed, this has been
the criterion for the most common classification scheme of these
baryons.  Table~\ref{tab:clasif} summarizes the quark content and quantum
numbers of the baryons considered in this study, classified so that
$S_h$ is well defined, and to which we shall refer as the
$S_h$-basis. Being ground state baryons,  a total orbital angular
momentum $L=0$ is assumed.

\begin{table}[h!!!]
\caption{ Quantum numbers and quark content of ground-state doubly
heavy baryons. }\label{tab:clasif}
\vspace{-3mm}
\footnotesize
\begin{tabular}{cccc||cccc}
\hline
Baryon   & Quark content 
                      & $S_h$ 
                           & $J^\pi$  &Baryon         & Quark content 
                                                                & $S_h$ 
                                                                   & $J^\pi$\vspace{-.1cm}\\ 
            &(l=u,d)    &   &         &                &           &   &         \\ \hline
$\Xi_{cc}$   & \{c~c\}~l & 1 & 1/2$^+$ & $\Omega_{cc}$   & \{c~c\}~s & 1 & 1/2$^+$ \\ 
$\Xi_{cc}^*$ & \{c~c\}~l & 1 & 3/2$^+$ & $\Omega_{cc}^*$ & \{c~c\}~s & 1 & 3/2$^+$ \\ 
$\Xi_{bb}$   & \{b~b\}~l & 1 & 1/2$^+$ & $\Omega_{bb}$   & \{b~b\}~s & 1 & 1/2$^+$ \\ 
$\Xi_{bb}^*$ & \{b~b\}~l & 1 & 3/2$^+$ & $\Omega_{bb}^*$ & \{b~b\}~s & 1 & 3/2$^+$ \\ 
$\Xi_{bc}$   & \{b~c\}~l & 1 & 1/2$^+$ & $\Omega_{bc}$   & \{b~c\}~s & 1 & 1/2$^+$ \\ 
$\Xi_{bc}^*$ & \{b~c\}~l & 1 & 3/2$^+$ & $\Omega_{bc}^*$ & \{b~c\}~s & 1 & 3/2$^+$ \\ 
$\Xi_{bc}'$  & [b~c]~l   & 0 & 1/2$^+$ & $\Omega_{bc}'$  & [b~c]~s   & 0 & 1/2$^+$ \\ 
\hline
\end{tabular}%
\end{table}
Due to the finite value of the heavy quark masses, the hyperfine
interaction between the light quark and any of the heavy quarks can
admix both $S_h=0$ and $S_h=1$ spin components into the wave
function. This mixing is
negligible for $cc$ and
$bb$ baryons, as the antisymmetry of the wave function would require higher
orbital angular momenta or radial excitations. On the other hand, in the $bc$ sector, one
would expect the role of the mixing to be noticeable and actual
physical states to be admixtures of the $B_{bc}$ and $B'_{bc}$ ($B$ =
$\Xi$, $\Omega$) states listed in Table~\ref{tab:clasif}.

Masses are rather insensitive to the mixing, and most calculations
simply ignore it and use the $S_h$-basis. Roberts and Pervin
\cite{pervin1} took the issue of the hyperfine mixing and its role in
semileptonic decays. They noticed that, working in the $S_h$-basis, the
decay width of $\Xi_{bc}$ ($\Omega_{bc}$) greatly differs from that of
$\Xi'_{bc}$ ($\Omega'_{bc}$)  so that mixing could be of relevance for these
processes.
In Ref.~\cite{pervin2} they found
that indeed mixing had  an enormous impact in 
 semileptonic decays of doubly-heavy baryons. 

We have studied the effect of this mixing in both semileptonic and
electromagnetic decays in Refs.~ \cite{albertus2} and \cite{albertus3}. Our
study on the semileptonic decay widths of $bc$ baryons qualitatively corroborate
the findings of Roberts and Pervin. 
In the  electromagnetic case 
 the decay widths
are  proportional to $(M_I-M_F)^3$, with $M_I,M_F$
the initial and final baryon  masses,  showing thus a strong 
dependence on the actual baryon masses.

\section{Results and discussion}

Table~\ref{tab:masses} shows our results for the masses of the unmixed
states. We compare them to the results from Refs.~ \cite{ebert1} and
\cite{pervin1}. Details on the model used can be found in Ref.~
\cite{albertus}.
\begin{table}
\caption{  Masses ({\rm in}\ MeV) for unmixed states.}\label{tab:masses}
\vspace{-3mm}
\footnotesize
\begin{tabular}{lccc||lccc}
\hline
& This work&\cite{ebert1}&\cite{pervin1} & &This work&\cite{ebert1}&\cite{pervin1}\\ \hline
$M_{\Xi_{cc}}$      &3613 &3620  & 3676    &$M_{\Omega_{cc}}$          & 3712 & 3778  &3815\\
$M_{\Xi_{cc}^*}$    &3707  &3727  &3753     &$M_{\Omega_{cc}^*}$        & 3795 & 3872  &3876 \\ 
$M_{ \Xi_{bb} }$    &10198 &10202 &10340    &$M_{\Omega_{bb}}$         & 10269 &10359 &10454\\ 
$M_{\Xi_{bb}^*}$    &10237 &10237 &10367    &$M_{\Omega_{bb}^*}$        & 10307 &10389 &10486\\ 
$M_{\Xi_{bc}}$      &6928  &6933  &7020     &$M_{\Omega_{bc}}$         & 7013  &7088  &7147 \\ 
$M_{\Xi_{bc}'}$     &6958  &6963  & 7044    &$M_{\Omega_{bc}'}$        & 7038  &7116  &7166\\
$M_{\Xi_{bc}^*}$    &6996  &6980  &7078     &$M_{\Omega_{bc}^*}$        & 7075  &7130  &7191\\ 
\hline
\end{tabular}%
\end{table}
Mixed ${bc}$ states are obtained by diagonalizing the corresponding
mass matrices. In our calculation, the mixed states and masses are
given by~\cite{albertus2}
\begin{eqnarray}
\label{eq:mix}
&&\hspace*{-.9cm}\Xi\,_{bc}^{(1)}=
\hspace{.3cm}0.902\,\Xi\,'_{bc}+0.431\,\Xi\,_{bc}\,,\ 
M_{\Xi\,_{bc}^{(1)}}=6967\,{\rm MeV},\nonumber\\
&&\hspace*{-.9cm}\Xi\,_{bc}^{(2)}= -0.431\,\Xi\,'_{bc}+0.902\,\Xi\,_{bc}\, ,\ 
M_{\Xi\,_{bc}^{(2)}}= 6919\,{\rm MeV},\nonumber\\
&&\hspace*{-.9cm}\Omega\,_{bc}^{(1)}=
\hspace{.25cm}0.899\,\Omega\,'_{bc}+0.437
\,\Omega\,_{bc}\,,\ 
M_{\Omega\,_{bc}^{(1)}}=7046\,{\rm MeV},\nonumber\\
&&\hspace*{-.9cm}\Omega\,_{bc}^{(2)}=
-0.437\,\Omega\,'_{bc}+0.899\,\Omega\,_{bc}\, ,\ 
M_{\Omega\,_{bc}^{(2)}}= 7005\,{\rm MeV}.
\end{eqnarray}
Comparing with Table~\ref{tab:masses}, we see small changes in the masses 
 when mixing is taken into account. However, as shown in 
Eq.~(\ref{eq:mix}), the admixture is important and it can affect the
decay widths.

The results for $b\to c$ semileptonic decays in the unmixed case 
are shown in Table~\ref{tab:widths}, where for
comparison we also show the results obtained in Refs.\cite{ebert2,faessler},
 within different relativistic approaches, and in the
nonrelativistic calculation of Ref.\cite{pervin2}. Our results are in
a global fair agreement with the ones in Ref.\cite{ebert2}.  As for
the other relativistic calculation in Ref.\cite{faessler}, the
agreement is fair for transitions with a $bc$ baryon in the initial
state but there is an approximate factor of 2 discrepancy for
transitions with a $bc$ baryon in the final state. The nonrelativistic
calculation in Ref.\cite{pervin2} also gives results that are roughly
a factor of 2 smaller than ours for all decays. A very interesting
feature of the decay widths shown in Table~\ref{tab:widths} is that
they are very different for transitions involving $\Xi_{bc}$ or
$\Xi'_{bc}$ ($\Omega_{bc}$ or $\Omega'_{bc}$). This means, as
suggested in Ref.\cite{pervin1}, that mixing in those states, provided
the admixture coefficients are large, can have a great impact on the
decay widths.
\begin{table}
\caption{  Semileptonic decay widths  $({\rm in\ units \ of}\ 10^{-14}\ {\rm GeV})$ for unmixed states.
We use $|V_{cb}|=0.0413$.  $l=e,\mu$.} \label{tab:widths}
\vspace{-3mm}
\footnotesize
\begin{tabular}[tH]{lcccc||lcccc}
\hline
&\hspace*{-.5cm} This
work\hspace*{-.25cm}&\cite{ebert2}&\hspace*{-.25cm}\cite{faessler}&\hspace*{-.25cm}\cite{pervin2}
&&\hspace*{-.25cm}This
work\hspace*{-.25cm}&\cite{ebert2}&\hspace*{-.25cm}\cite{faessler}&\hspace*{-.25cm}\cite{pervin2}\\\hline
$\Gamma(\Xi_{bb}^*\to\Xi_{bc}'\,l\bar\nu_l)$ &  $1.08$  &$0.82$&
\hspace*{-.25cm}$0.36$&\hspace*{-.25cm}--
&$\Gamma(\Omega_{bb}^*\to\Omega_{bc}'\,l\bar\nu_l)$ & $1.14$
&$0.85$&\hspace*{-.25cm}$0.42$&\hspace*{-.25cm}-- \\ 
$\Gamma(\Xi_{bb}^*\to\Xi_{bc}\,l\bar\nu_l)$
&$0.36$&$0.28$&\hspace*{-.25cm}$0.14$&\hspace*{-.25cm}--
&$\Gamma(\Omega_{bb}^*\to\Omega_{bc}\,l\bar\nu_l)$
&$0.38$&$0.29$&\hspace*{-.25cm}$0.15$&\hspace*{-.25cm}--\\ 
$\Gamma(\Xi_{bb}\to\Xi_{bc}'\,l\bar\nu_l)$ &  $1.09$ &$0.82$&\hspace*{-.25cm}$0.43$&
\hspace*{-.25cm}$0.41$
&$\Gamma(\Omega_{bb}\to\Omega_{bc}'\,l\bar\nu_l)$ & $1.16$
&$0.83$&\hspace*{-.25cm}$0.48$&\hspace*{-.25cm}$0.51$ \\ 
$\Gamma(\Xi_{bb}\to\Xi_{bc}\,l\bar\nu_l)$ 
&$2.00$&$1.63$&\hspace*{-.25cm}$0.80$&\hspace*{-.25cm}$0.69$
&$\Gamma(\Omega_{bb}\to\Omega_{bc}\,l\bar\nu_l)$  &
$2.15$&$1.70$&\hspace*{-.25cm}$0.86$&\hspace*{-.25cm}$0.92$\\ 
$\Gamma(\Xi_{bc}'\to\Xi_{cc}\,l\bar\nu_l)$ & 
$1.36$&$0.88$&\hspace*{-.25cm}$1.10$&\hspace*{-.25cm}--
&$\Gamma(\Omega_{bc}'\to\Omega_{cc}\,l\bar\nu_l)$ &
$1.36$&$0.95$&\hspace*{-.25cm}$0.98$&\hspace*{-.25cm}-- \\ 
$\Gamma(\Xi_{bc}\to\Xi_{cc}\,l\bar\nu_l)$  &$ 2.57
$&$2.30$&\hspace*{-.25cm}$2.10$&\hspace*{-.25cm}$1.38$
&$\Gamma(\Omega_{bc}\to\Omega_{cc}\,l\bar\nu_l)$  &
$2.58$&$2.48$&\hspace*{-.25cm}$1.88$&\hspace*{-.25cm}$1.54$\\ 
$\Gamma(\Xi_{bc}'\to\Xi_{cc}^*\,l\bar\nu_l)$ & 
$2.35$&$1.70$&\hspace*{-.25cm}$2.01$&\hspace*{-.25cm}-- 
&$\Gamma(\Omega_{bc}'\to\Omega_{cc}^*\,l\bar\nu_l)$ & $2.35$
&$1.83$&\hspace*{-.25cm}$1.93$&\hspace*{-.25cm}--\\ 
$\Gamma(\Xi_{bc}\to\Xi_{cc}^*\,l\bar\nu_l)$  &$ 0.75 $ &$0.72$&\hspace*{-.25cm}$0.64$&\hspace*{-.25cm}$0.52$
&$\Gamma(\Omega_{bc}\to\Omega_{cc}^*\,l\bar\nu_l)$ 
&$0.76$&$0.74$&\hspace*{-.25cm}$0.62$&\hspace*{-.25cm}$0.56$\\
\hline
\end{tabular}%
\end{table}

In Table \ref{tab:emunmixed} we show our results for electromagnetic
decays in the unmixed case.  Branz et al.~\cite{branz} have also studied 
electromagnetic
decays of doubly heavy baryons within a relativistic constituent quark
model. The agreement with our results is very poor in this case in part due
to the different masses used in both calculations. The agreement
improves in most cases if we  divide out  the
$(M_I-M_F)^3$ mass factor discussed above. Still the  differences 
are in the range 50-80\%. This can be seen in
Table~\ref{tab:gunmixeddiv}.
\begin{table}
\caption{Electromagnetic decay widths $({\rm in\ units\ of}\ 10^{-8}\ 
{\rm GeV})$ for unmixed states. }
\label{tab:emunmixed}
\begin{tabular}{lcc||lcc}
\hline
                                 & This work   & \cite{branz}  &      & This work    & \cite{branz}  \\
$\Xi_{bcu}^*\to\Xi'_{bcu}\gamma$   &  $4.04$    & $0.28\pm0.01$  & $\Omega_{bcs}^*\to\Omega'_{bcs}\gamma$    &  $3.69$  & $0.16\pm 0.01$   \\
$\Xi_{bcd}^*\to\Xi'_{bcd}\gamma$   &  $4.04$     & $0.28\pm0.01$ &                                       &             \\\hline
$\Xi_{bcu}^*\to\Xi_{bcu}\gamma$    &  $105$      & $49\pm9$ & $\Omega_{bcs}^*\to\Omega_{bcs}\gamma$     &  $20.9$  & $0.12\pm 0.02$  \\
$\Xi_{bcd}^*\to\Xi_{bcd}\gamma$    &  $50.5$     & $24\pm4 $    &                              &            \\\hline
$\Xi_{bcu}'\to\Xi_{bcu}\gamma$     &  $0.992$   & $1.56\pm0.08$  & $\Omega_{bcs}'\to\Omega_{bcs}\gamma$      &  $0.568$ & $1.26\pm5$  \\
$\Xi_{bcd}'\to\Xi_{bcd}\gamma$     &  $0.992$    & $1.56\pm0.08$ &                                        &            \\\hline
\end{tabular}
\end{table}

\begin{table}\caption{Electromagnetic decay widths, divided by $(M_I-M_F)^3$, 
(in units of  $(10^{-5}\ {\rm GeV^{-2}})$ for unmixed states.}
\label{tab:gunmixeddiv}
\begin{tabular}{lcc||lcc}
\hline
                                 & This work & Branz {\it et al.}    & 
                                 & This work &  Branz {\it et al.} \\ 

\hline
$\Xi_{bcu}^*\to\Xi'_{bcu}\gamma$   &  $73.6$    &     $57.0$         
& $\Omega_{bcs}^*\to\Omega'_{bcs}\gamma$    &  $72.8$   &   $58.3$   \\
$\Xi_{bcd}^*\to\Xi'_{bcd}\gamma$   &  $73.6$    &     $57.0$         &  
                                       &           &            \\\hline
$\Xi_{bcu}^*\to\Xi_{bcu}\gamma$    &  $333.9$   &       $471$       
 & $\Omega_{bcs}^*\to\Omega_{bcs}\gamma$     &  $87.7$   &   $160$    \\
$\Xi_{bcd}^*\to\Xi_{bcd}\gamma$    &  $160.6$   &       $231$    
    &                                         &           &            \\\hline
$\Xi_{bcu}'\to\Xi_{bcu}\gamma$     &  $36.7$    &       $57.8$  
     & $\Omega_{bcs}'\to\Omega_{bcs}\gamma$      &  $36.3$ & $57.4$    \\
$\Xi_{bcd}'\to\Xi_{bcd}\gamma$     &  $36.7$    &       $57.8$       & 
                                        &           &            \\\hline
\end{tabular}
\end{table}

\begin{table}
\caption{  Semileptonic decay widths  $({\rm in\ units\ of}\ 10^{-14}\ 
{\rm GeV})$ for mixed states.
}\label{tab:newwidths}
\vspace{-3mm}
\footnotesize
\begin{tabular}{lcc||lcc}
\hline
  &This work&\cite{pervin2}&&This work&\cite{pervin2}\\\hline
$\Gamma(\Xi_{bb}^*\to\Xi^{(1)}_{bc}\,l\bar\nu_l)$ &  $0.47$& --  
&$\Gamma(\Omega_{bb}^*\to\Omega^{(1)}_{bc}\,l\bar\nu_l)$ & $0.48$&--  \\ 
$\Gamma(\Xi_{bb}^*\to\Xi^{(2)}_{bc}\,l\bar\nu_l)$\ &$0.99$&--
&$\Gamma(\Omega_{bb}^*\to\Omega^{(2)}_{bc}\,l\bar\nu_l)$\ &$1.06$&--\\ 
$\Gamma(\Xi_{bb}\to\Xi^{(1)}_{bc}\,l\bar\nu_l)$ &  $2.21$& $0.95$ 
&$\Gamma(\Omega_{bb}\to\Omega^{(1)}_{bc}\,l\bar\nu_l)$ & $2.36$& $0.99$  \\ 
$\Gamma(\Xi_{bb}\to\Xi^{(2)}_{bc}\,l\bar\nu_l)$  &$0.85$& $0.33$
&$\Gamma(\Omega_{bb}\to\Omega^{(2)}_{bc}\,l\bar\nu_l)$  & $0.91$& $0.30$\\ 
$\Gamma(\Xi^{(1)}_{bc}\to\Xi_{cc}\,l\bar\nu_l)$ &  $0.38$& --
&$\Gamma(\Omega^{(1)}_{bc}\to\Omega_{cc}\,l\bar\nu_l)$ & $0.37$& -- \\ 
$\Gamma(\Xi^{(2)}_{bc}\to\Xi_{cc}\,l\bar\nu_l)$  &$ 3.50$&$ 1.92 $
&$\Gamma(\Omega^{(2)}_{bc}\to\Omega_{cc}\,l\bar\nu_l)$  & $3.52$& $1.99$\\ 
$\Gamma(\Xi^{(1)}_{bc}\to\Xi_{cc}^*\,l\bar\nu_l)$ &  $3.14$&--
&$\Gamma(\Omega^{(1)}_{bc}\to\Omega_{cc}^*\,l\bar\nu_l)$ & $3.14$&-- \\ 
$\Gamma(\Xi^{(2)}_{bc}\to\Xi_{cc}^*\,l\bar\nu_l)$  &$0.017$& $0.026$
&$\Gamma(\Omega^{(2)}_{bc}\to\Omega_{cc}^*\,l\bar\nu_l)$  &$0.014$&$0.013$\\
\hline
\end{tabular}%
\end{table}

$b\to c$ semileptonic decay widths involving the mixed states
$\Xi\,_{bc}^{(1)},\,\Xi\,_{bc}^{(2)}$ and
$\Omega\,_{bc}^{(1)},\,\Omega\,_{bc}^{(2)}$ are now given in
Table~\ref{tab:newwidths}. We see rather big changes from the values
in Table~\ref{tab:widths} where unmixed states were used.  Special
attention deserves the $B^{(2)}_{bc}\to B^*_{cc}$ transitions where
the width reduces by a large factor of $44$ (54) for the
$\Xi^{(2)}_{bc}\to\Xi^*_{cc}$ ($\Omega^{(2)}_{bc}\to \Omega^*_{cc}$)
decay compared to the unmixed case. This can be easily understood by
taking into account that
$B^{(2)}_{bc}\approx\left(|qc;0\rangle\otimes|b;\frac12\rangle\right)^{J=1/2}$. In
the latter state the light and $c$ quarks are coupled to spin 0,
whereas in the $B^*_{cc}$ the light and any of the $c$ quarks are in a
relative spin 1 state. In any spectator calculation, as the ones here
and in Ref.\cite{pervin2}, the amplitude for the
$\left(|qc;0\rangle\otimes|b;\frac12\rangle\right)^{J=1/2}\to
B^*_{cc}$ transition cancels due to the orthogonality of the different
spin states of the spectator quarks in the initial and final
baryons. The fact that $B^{(2)}_{bc}$ slightly deviates from$
\left(|qc;0\rangle\otimes|b;\frac12\rangle\right)^{J=1/2}$ produces a
non zero, but small, decay width.
\begin{table}
\caption{Electromagnetic decay widths~$({\rm in\ units\ of}\ 10^{-8}\ 
{\rm GeV})$ for mixed states.}
\label{tab:emmixed}
\begin{tabular}{lcc||lcc}
\hline
                                         & This work   & \cite{branz}   && This work & \cite{branz} \\\hline
$\Xi_{bcu}^*\to\Xi^{(1)}_{bcu}\gamma$       &  $6.05$  & $0.15\pm0.02$   & $\Omega_{bcs}^*\to\Omega^{(1)}_{bcs}\gamma$    &  $0.31$ & $(1\pm1)\cdot10^{-4}$   \\
$\Xi_{bcd}^*\to\Xi^{(1)}_{bcd}\gamma$       &  $0.12$  & $(2\pm2)\cdot 10^{-4}$  &                                             &             \\\hline
$\Xi_{bcu}^*\to\Xi^{(2)}_{bcu}\gamma$       &  $73.9$  & $46\pm10$   & $\Omega_{bcs}^*\to\Omega^{(2)}_{bcs}\gamma$    &  $50.2$ &$29\pm3$ \\
$\Xi_{bcd}^*\to\Xi^{(2)}_{bcd}\gamma$       &  $103$   &$51\pm6$    &                                             &            \\\hline
$\Xi^{(1)}_{bcu}\to\Xi^{(2)}_{bcu}\gamma$    &  $12.4$  &$14\pm3$   & $\Omega_{bcs}^{(1)}\to\Omega^{(2)}_{bcs}\gamma$ &  $8.52$ &$21\pm2$   \\
$\Xi^{(1)}_{bcd}\to\Xi^{(2)}_{bcd}\gamma$    &  $20.9$  &$31\pm4$   &                                             &            \\\hline
\end{tabular}
\end{table}

Our results for the electromagnetic decay width for mixed states are
enclosed in Table~\ref{tab:emmixed}. To the best of our knowledge ours
was the first calculation which took into account the effect of the
mixing in the electromagnetic decay width of $bc$-baryons. As for the
unmixed case, we find a better qualitative agreement with the results
of Ref.~\cite{branz} when we show the decay widths divided by the
$(M_I-M_F)^3$ mass factor, see Table~\ref{tab:gdiv}.

\begin{table}
\caption{Electromagnetic decay widths, divided by $(M_I-M_F)^3$ in units of 
 $(10^{-5}\ {\rm GeV^{-2}})$ for mixed states.}\label{tab:gdiv}
\begin{tabular}{lcc||lcc}
\hline
                                        & This work & Branz {\it et al.}    &                                             & This work &  {\it Branz et al.} \\ 
\hline
$\Xi_{bcu}^*\to\Xi^{(1)}_{bcu}\gamma$       &  $248$  &   $293$              & $\Omega_{bcs}^*\to\Omega^{(1)}_{bcs}\gamma$    &  $12.7$   &  $0.8$     \\
$\Xi_{bcd}^*\to\Xi^{(1)}_{bcd}\gamma$       &  $4.9$  &   $0.39$             &                                             &          &             \\\hline
$\Xi_{bcu}^*\to\Xi^{(2)}_{bcu}\gamma$       &  $161.8$ &  $261$              & $\Omega_{bcs}^*\to\Omega^{(2)}_{bcs}\gamma$    &  $146$   &   $218$    \\
$\Xi_{bcd}^*\to\Xi^{(2)}_{bcd}\gamma$       &  $226$   &  $290$              &                                             &          &             \\\hline
$\Xi^{(1)}_{bcu}\to\Xi^{(2)}_{bcu}\gamma$    &  $112$   &  $126$              & $\Omega_{bcs}^{(1)}\to\Omega^{(2)}_{bcs}\gamma$ &  $124$   &   $215$     \\
$\Xi^{(1)}_{bcd}\to\Xi^{(2)}_{bcd}\gamma$    &  $189$   &  $280$              &                                             &          &             \\\hline
\end{tabular}
\end{table}

\section{Conclusions}
We qualitatively confirm the findings in Refs.~\cite{pervin1,pervin2}
as to the relevance of hyperfine mixing in $b\to c$ semileptonic
decays of doubly heavy baryons. Actual results differ by a factor of
two. We find mixing is also very important for electromagnetic decays.
In this latter case decay widths are very sensitive to the actual
baryon masses We have compared our results for the electromagnetic
decay widths with those from Ref.~\cite{branz}. Predictions are rather
different, although a better agreement is found if the dependence of the witdh
on $(M_I-M_F)^3$
 is divided out.  Electromagnetic decay studies demand an
accurate determination of the masses.
 

\bibliographystyle{aipproc}

\end{document}